\documentclass[12pt,fleqn]{article}

\usepackage{amsmath,amssymb}
\usepackage{graphicx}

\def\eqref#1{(\ref{#1})}

\begin{document}
\title{
{\sf QCD Sum-Rule Bounds on the Light Quark Masses}
}
\author{
 T.G.~Steele\thanks{email: Tom.Steele@usask.ca}\\
  \textsl{Department of Physics and Engineering Physics}\\
  \textsl{University of Saskatchewan}\\
  \textsl{Saskatoon, Saskatchewan, S7N 5E2}\\
  \textsl{Canada}
}
\maketitle

\begin{abstract}
QCD sum-rules are related to an integral of a hadronic
spectral function, and hence must satisfy integral inequalities which
follow from positivity of the spectral function. Development of these
H\"older inequalities and their application to the Laplace sum-rule for pions
lead to a lower bound on the average of the non-strange  $2\,{\rm GeV}$ light-quark masses  
 in the $\overline{\rm MS}$ scheme.
\end{abstract}
The light quark masses are  fundamental parameters of QCD, and 
determination of their values is of importance for high-precision 
QCD phenomenology and lattice simulations involving dynamical quarks.
In this paper the development of H\"older inequalities for QCD Laplace sum-rules \cite{sr_holder} 
is briefly reviewed.  These  techniques are then used to obtain bounds on the 
non-strange (current) quark masses $m_n=\left(m_u+m_d\right)/2$  evaluated at $2\,{\rm GeV}$ in the $\overline{\rm MS}$ scheme, updating and
extending the H\"older inequality results of ref.\ \cite{holder_bounds}.

Although it is possible to obtain quark mass ratios in various contexts \cite{mass_ratios}, the only methods
which have been able to determine the {\em absolute} non-strange quark mass scales are the 
lattice (see \cite{lattice_mass} for recent results with two dynamical flavours)  
and QCD sum-rules 
\cite{holder_bounds,BNRY,bounds,other_bounds,sumrule_mass,3pi}.\footnote{An overview of selected lattice and sum-rule results for both non-strange and strange masses can be found  in \protect\cite{gupta}.}

In sum-rule and lattice  approaches, the pseudoscalar or scalar channels are used since they have the 
strongest dependence on the quark masses.  
This is exemplified by the correlation function 
 $\Pi_5\left(Q^2\right)$  of 
renormalization-group (RG) invariant pseudoscalar
 currents with quantum numbers of the pion:
\begin{eqnarray}
\Pi_{5}\left(Q^2\right)&=&i\int {\rm d}^4x\, {\rm e}^{iq\cdot x}\left\langle O \vert T\left[ J_{5}(x) 
J_5(0)\right] \vert O \right\rangle
\label{corr_fn}\\
J_5(x)&=&\frac{1}{\sqrt{2}}\left(m_u+m_d\right)\left[\bar u(x)i\gamma_5u(x)-
\bar d(x)i\gamma_5d(x)\right]\quad .
\label{current}
\end{eqnarray} 
The Laplace sum-rule  is obtained by
 by applying the 
Borel transform operator \cite{SVZ} $\hat B$ 
\begin{equation}
\hat B\equiv 
\lim_{\stackrel{N,~Q^2\rightarrow \infty}{Q^2/N\equiv M^2}}
\frac{\left(-Q^2\right)^N}{\Gamma(N)}\left(\frac{{\rm d}}{{\rm d}Q^2}\right)^N
\label{borel_op}
\end{equation}  
to the dispersion relation for $\Pi_5\left(Q^2\right)$
\begin{equation}
\Pi_{5}\left(Q^2\right)=a+b Q^2
+\frac{Q^4}{\pi}\int\limits_{4m_\pi^2}^\infty
\frac{ \rho_{_5}(t)}{t^2\left(t+Q^2\right)}\, {\rm d}t\quad ,
\label{dispersion}
\end{equation}
where $\rho_{_5}(t)$ is the hadronic spectral function appropriate to the pion quantum numbers, and
the quantities $a$ and $b$ represent subtraction constants. 
The resulting Laplace sum-rule relating  the theoretically-determined 
quantity
\begin{equation}
R_{5}\left(M^2\right)=M^2\hat B\left[\Pi_{5}\left(Q^2\right)\right]
\label{R5_def}
\end{equation}
 to phenomenology is
\begin{equation}
R_5\left(M^2\right)=\frac{1}{\pi}\int\limits_{4m_\pi^2}^\infty \rho_{_5}(t) \exp{\left(-\frac{t}{M^2}\right)}\,{\rm d}t \quad .
\label{basic_sr}
\end{equation}

  Perturbative contributions to  $R_5\left(M^2\right)$ are known up to 
four-loop order in the $\overline{\rm MS}$ \cite{BNRY,chetyrkin}.  
Infinite correlation-length vacuum effects in  $R_5\left(M^2\right)$
are represented by  the (non-perturbative) QCD condensate contributions \cite{BNRY,SVZ,Bagan}.  
In addition to the QCD condensate contributions the pseudoscalar (and scalar) correlation functions
are sensitive to finite correlation-length vacuum effects described by direct instantons \cite{dorokhov}
 in the instanton 
liquid model \cite{EVS}.  Combining all these results,  
the total result for $R_5\left(M^2\right)$ to leading order 
in the light-quark masses is \cite{holder_bounds}
\begin{eqnarray}
{ R}_5\left(M^2\right)&=&\frac{3m_n^2M^4}{8\pi^2}\left(   
1+4.821098 \frac{\alpha}{\pi}+21.97646\left(\frac{\alpha}{\pi}\right)^2+53.14179\left(\frac{\alpha}{\pi}\right)^3
\right)
\nonumber\\
& &+m_n^2\left(
-\langle m\bar q q\rangle 
+\frac{1}{8\pi}\langle \alpha G^2\rangle
+\frac{\pi\langle{\cal O}_6\rangle}{4M^2}
\right)
\nonumber\\
& &+m_n^2
\frac{3\rho_c^2M^6}{8 \pi^2} \mathrm{e}^{-\rho_c^2M^2/2 }
\left[   
  K_0\left( {\rho_c^2M^2/2} \right) +
       K_1\left( {\rho_c^2M^2/2} \right)
\right]\quad ,
\label{R_5}
\end{eqnarray} 
where $\alpha$ and  $m_n=\left(m_u+m_d\right)/2$ are the $\overline{{\rm MS}}$ 
running coupling and quark masses at the scale $M$, and $\rho_c=1/(600\,{\rm MeV})$ represents the instanton size in the 
instanton liquid model \cite{EVS}. 
$SU(2)$ symmetry has been used for the dimension-four quark condensates ({\it i.e.} 
$(m_u+m_d)\langle \bar u u\rangle+ \bar d d\rangle)\equiv 4 m\langle \bar q q\rangle$),
and   $\langle {\cal O}_6\rangle$ denotes the dimension six quark condensates 
\begin{eqnarray}
\langle{\cal O}_6\rangle&\equiv& \alpha_s
\biggl[ 
\left(2\langle \bar u \sigma_{\mu\nu}\gamma_5
T^au\bar u \sigma^{\mu\nu}\gamma_5T^a u\rangle
+ u\rightarrow d\right)
 -4\langle \bar u \sigma_{\mu\nu}\gamma_5T^au\bar d 
 \sigma^{\mu\nu}\gamma_5 T^a d\rangle
\biggr. \nonumber\\
& &\qquad\qquad 
\biggl.+\frac{2}{3}
\langle \left(   
\bar u \gamma_\mu T^a u+\bar d \gamma_\mu T^a d 
\right)
\sum_{u,d,s}\bar q \gamma^\mu T^aq
\rangle\biggr]
\label{o6}
\end{eqnarray}
The vacuum saturation hypothesis \cite{SVZ} will be used as a reference value for $\langle{\cal O}_6\rangle$
\begin{equation}
\langle{\cal O}_6\rangle=f_{vs}\frac{448}{27}\alpha
\langle \bar q q\bar q q\rangle
=f_{vs}3\times 10^{-3} {\rm GeV}^6 
\label{o61}
\end{equation}
where $f_{vs}=1$ for exact vacuum saturation.  Larger values 
of effective dimension-six
operators found in \cite{dimsix1,dimsix2} imply that $f_{vs}$  could be as 
large as $f_{vs}=2$.
The quark condensate is determined by the GMOR (PCAC) relation 
\begin{equation}
\left(m_u+m_d\right)\langle \bar u u+\bar d d\rangle=4m\langle \bar q q\rangle=
-2f_\pi^2m_\pi^2
\label{GMOR}
\end{equation}
where $f_\pi=93\,{\rm MeV}$.  A recent determination
of the gluon condensate $\langle \alpha G^2\rangle$ will be used:
\cite{narison2}
\begin{equation}
\langle \alpha G^2\rangle=\left(0.07\pm 0.01\right)\,{\rm GeV^4}\quad .
\label{aGG}
\end{equation}
However, it should be noted that there is some discrepancy between 
\cite{narison2} 
 and the smaller value $\langle\alpha G^2\rangle=\left( 0.047\pm 0.014\right)\,{\rm GeV^4}$ found in \cite{dimsix2}.

Note that {\em all }the theoretical contributions in \eqref{R_5} are proportional to  $m_n^2$, demonstrating that the quark mass sets
the scale of the pseudoscalar channel.  This dependence on the quark mass can be singled out as follows:
\begin{equation}
{ R}_5\left(M^2\right)=\left[m_n(M)\right]^2\,G_5\left(M^2\right)
\label{G_def}
\end{equation}
where $G_5$ is independent of $m_n$ and is trivially extractable from \eqref{R_5}.
Higher-loop perturbative contributions in (\ref{R_5}) are thus significant since they can 
effectively enhance the quark mass with increasing loop order.  

Determinations of the non-strange quark mass $m_n$ using the sum-rule \eqref{basic_sr} require input of a phenomenological model for the spectral function $\rho_{_5}(t)$.  The mass $m_n$ can then be determined by fitting 
to find the best agreement between the phenomenological model and the theoretical prediction 
respectively appearing on the right- and left-hand sides of \eqref{basic_sr}.  For example, the simple resonance(s) plus continuum model
\begin{equation}
\frac{1}{\pi}\rho_{_5}(t)=2f_\pi^2m_\pi^4\left[
\delta\left(t-m_\pi^2\right)+\frac{F_\Pi^2 M_\Pi^4}{f_\pi^2m_\pi^4}\delta\left(t-M_\Pi^2\right)
\right]
+\Theta\left(t-s_0\right)\frac{1}{\pi}\rho^{QCD}(t)
\label{res_model}
\end{equation}
represents the pion pole ($m_\pi$), a narrow-width approximation to the pion excitation ($M_\Pi$) 
such as the $\Pi(1300)$, and a QCD continuum above the continuum threshold $t=s_0$. Of course more 
detailed phenomenological models can be considered which take into account possible width effects for the pion 
excitation, further resonances, resonance(s) enhancement of the $3\pi$ continuum {\it etc.} This leads to significant 
model dependence which partially accounts for the spread of theoretical estimates in \cite{sumrule_mass,3pi}.
Since the common phenomenological portion of all these models is the pion pole, it is valuable to extract quark mass
{\em bounds} which only rely upon the input of the pion pole on the phenomenological side of \eqref{basic_sr}.

The existence of such bounds is easily seen by separating the pion pole out from $\rho_{_5}(t)$, in which case 
\eqref{basic_sr} becomes
\begin{equation}
\left[m_n(M)\right]^2 G_5\left(M^2\right)=2f_\pi^2m_\pi^4+
\frac{1}{\pi}\int\limits_{9m_\pi^2}^\infty \rho_{_5}(t) \exp{\left(-\frac{t}{M^2}\right)}\,{\rm d}t \quad .
\label{pion_pole_sr}
\end{equation}
Since $\rho_{_5}(t)\ge 0$ in the integral appearing on the right-hand side of \eqref{pion_pole_sr}, a bound on the
quark mass is obtained:
\begin{equation}
m_n(M)\ge\sqrt{\frac{2f_\pi^2m_\pi^4}{G_5\left(M^2\right)}}\quad .
\label{simple_bound}
\end{equation}
Analysis of these bounds following from simple positivity of the ``residual'' portion on the right-hand side of \eqref{pion_pole_sr} was studied in \cite{BNRY,bounds}.

Improvements upon the positivity bound of \eqref{simple_bound} are achieved by developing more stringent inequalities 
based on the positivity of $\rho_{_5}(t)$.
Since $\rho_{_5}(t)\ge 0$, the right-hand (phenomenological) side of 
(\ref{basic_sr}) must satisfy integral inequalities over a measure  ${\rm d}\mu=\rho_{_5}(t)\,{\rm d}t$.
In particular, H\"older's inequality over a measure ${\rm d}\mu$ is 
\begin{equation}
\biggl|\int_{t_1}^{t_2} f(t)g(t) {\rm d}\mu \biggr|\! \le \!
\left(\int_{t_1}^{t_2} \big|f(t)\big|^ p {\rm d}\mu \right)^{\frac{1}{p}}
\!\!\!\left(\int_{t_1}^{t_2} \big|g(t)\big|^q {\rm d}\mu \right)^{\frac{1}{q}}
~,~
\frac{1}{p}+\frac{1}{q} =1~;~ p,~q\ge 1 \quad ,
\label{holder_ineq}
\end{equation}
which for  $p=q=2$   reduces to the familiar Schwarz inequality, implying 
that the H\"older inequality is a more general constraint. The H\"older inequality can be applied to 
Laplace sum-rules by identifying ${\rm d}\mu=\rho_{_5}(t)\,{\rm d}t$, $\tau=1/M^2$ and defining
\begin{equation}
S_5\left(\tau\right)=\frac{1}{\pi}\int\limits_{\mu_{th}}^\infty \!\!\rho_{_5}(t) e^{-t\tau}\,{\rm d}t
\label{s5}
\end{equation}
where $\mu_{th}$ will later be identified with $9m_\pi^2$.  Suitable choices of 
$f(t)$ and $g(t)$ in the H\"older inequality (\ref{holder_ineq}) yield the following 
inequality  for $S_5(t)$ \cite{sr_holder}: 
\begin{equation}
S_5\left(\tau+(1-\omega)\delta\tau\right)\le \left[S_5\left(\tau\right)\right]^\omega
\left[S_5\left(\tau+\delta\tau\right)\right]^{1-\omega} 
\quad ,~ \forall~ 0\le \omega\le 1\quad .
\label{s5_ineq}
\end{equation}
In practical applications of this inequality,
$\delta\tau\le 0.1\,{\rm GeV^{-2}}$  is used,  in which case this inequality analysis becomes local (depending only on the 
Borel scale $M$ and not on $\delta\tau$) \cite{sr_holder,holder_bounds}.

To employ the H\"older inequality (\ref{s5_ineq}) we separate out the pion pole by setting $\mu_{th}=9m_\pi^2$ 
in (\ref{s5}).
\begin{equation}
S_5\left(M^2\right)=R_5\left(M^2\right)-2f_\pi^2m_\pi^4=\int\limits_{9m_\pi^2}^\infty \rho_{_5}(t)
 e^{-t\tau}\,{\rm d}t
\label{s5_fin}
\end{equation}
which has a right-hand side  in the standard form \eqref{s5} for applying the H\"older inequality. Note that simple positivity of $\rho_{_5}(t)$ gives the inequality
\begin{equation}
S_5\left(M^2\right)\ge 0
\label{positivity}
\end{equation}
which simply rephrases \eqref{simple_bound}.
Lower bounds on the quark mass $m_n$ can now be obtained by finding the minimum value of $m_n$ for which the
H\"older inequality (\ref{s5_ineq}) is satisfied. 
Introducing further phenomenological 
contributions ({\it e.g.} three-pion continuum)  give a slightly larger mass bound as will be discussed later.  
However,  if only the  pion pole is separated out, then the analysis is not subject to uncertainties introduced by the
 phenomenological model.

Although the details are still a matter of dispute, the overall 
validity of QCD predictions at the tau mass is evidenced by the 
analysis of the tau hadronic width, hadronic contributions to $\alpha_{EM}\left(M_Z\right)$ and the 
muon anomalous magnetic moment \cite{braaten_davier}, so 
we impose the inequality (\ref{s5_ineq}) at the tau mass scale  $M=M_\tau=1.77\,{\rm GeV}$. 
This also has the advantage of minimizing perturbative uncertainties in the running of $\alpha$ and $m_n$, since the
the PDG reference scale for the light-quark masses is at $2\,{\rm GeV}$ \cite{PDG}, in close proximity to $M_\tau$, and 
the result  $\alpha_s\left(M_\tau\right)= 0.33 \pm 0.02$ 
\cite{aleph} can thus be used to its maximum advantage. For the remaining small energy range in which the running of $\alpha$ and $m_n$ is needed, the four-loop $\beta$-function \cite{beta} and four-loop anomalous mass dimension \cite{gamma} 
with three active flavours are used, appropriate to the analysis of \cite{aleph}.  This use of the $2\,{\rm GeV}$ reference scale for $m_n$ 
combined with input of $\alpha\left(M_\tau\right)$ improves upon the perturbative uncertainties in 
\cite{holder_bounds} which employed $\alpha\left(M_Z\right)$ and a $1\,{\rm GeV}$ $m_n$ reference scale which 
necessitated matching through the (uncertain) $b$ and $c$ flavour thresholds.

Further theoretical uncertainties devolve from the QCD condensates
as given in \eqref{aGG} and \eqref{o61} with $1\le f_{vs}\le 2$, along with a 15\% uncertainty in the
instanton liquid parameter $\rho_c$ \cite{EVS}. The effect of higher-loop perturbative contributions to $R_5\left(M^2\right)$ on the
resulting $m_n$ bounds is estimated using an asymptotically-improved Pad\'e estimate \cite{apap} of the five-loop term, introducing a $138\left(\alpha/\pi\right)^4$ correction into \eqref{R_5}. Finally, we allow for the 
possibility that the overall scale of the instanton is 50\% uncertain.

The resulting H\"older inequality bound on the  $2.0\,{\rm GeV}$ $\overline{\rm MS}$ quark masses, updating the
analysis of \cite{holder_bounds},  is
\begin{equation}
m_n(2\,{\rm GeV})=\frac{1}{2}\left[ m_u(2\,{\rm GeV})+m_d(2\,{\rm GeV})\right]\ge 2.1\,{\rm MeV}
\label{final_pdg_bound}
\end{equation}
This final result is identical to previous bounds on  $m_n(1\,{\rm GeV})$ \cite{holder_bounds} after conversion to 
$2\,{\rm GeV}$ by the PDG \cite{PDG}, indicative of the consistency of perturbative inputs used in the two analyses.
The theoretical uncertainties in the quark mass bound (\ref{final_pdg_bound}) from the QCD parameters and 
(estimated) higher-order perturbative effects are less than 10\%, and the result (\ref{final_pdg_bound}) 
is the absolute lowest bound resulting from  the  uncertainty analysis.  The dominant sources of uncertainty are $\alpha\left(M_\tau\right)$ and potential higher-loop corrections.  The instanton size $\rho_c$ is the major source of non-perturbative uncertainty, but its effect is smaller than the perturbative sources of uncertainty.

Compared with the
positivity inequality  \eqref{positivity}, as first used to obtain quark mass bounds from QCD 
sum-rules \cite{BNRY,bounds},
the H\"older inequality leads to  quark mass bounds 50\% larger   for
identical theoretical and phenomenological inputs at $M=M_\tau$, demonstrating that the H\"older inequality 
provides  stringent constraints on the quark mass.  

Finally, we discuss the effects of extending the resonance model to include the $3\pi$ continuum calculated using lowest-order chiral perturbation theory \cite{3pi}  
\begin{eqnarray}
\frac{1}{\pi}\rho_{_5}(t)&=&2f_\pi^2m_\pi^4\left[
\delta\left(t-m_\pi^2\right)+\Theta\left(t-9m_\pi^2\right)\rho_{3\pi}(t)\frac{t}{18\left(16\pi^2f_\pi^2\right)^2}\right]
\\
\rho_{3\pi}(t)&=&\int\limits_{4m_\pi^2}^{\left(\sqrt{t}-m_\pi\right)^2}
\frac{{\rm d}u}{t}\sqrt{\lambda\left(1,\frac{u}{t},\frac{m_\pi^2}{t}\right)}\sqrt{1-\frac{4m_\pi^2}{u}}
\Biggl\{
5+\Biggr.
\nonumber\\
& &+\frac{1}{2\left(t-m_\pi^2\right)^2}\left[\frac{4}{3}\left(t-3\left(u-m_\pi^2\right)\right)^2
+\frac{8}{3}\lambda\left(t,u,m_\pi^2\right)\left(1-\frac{4m_\pi^2}{u}\right)+10m_\pi^4
\right]
\nonumber\\
& &\Biggl.+\frac{1}{t-m_\pi^2}\left[3\left(u-m_\pi^2\right)-t+10m_\pi^2
\right]\Biggr\}
\label{3pi_cont}
\\
\lambda(x,y,z)&=&x^2+y^2+z^2-2xy-2yz-2xz
\end{eqnarray}
which becomes $\rho_{3\pi}(t)\rightarrow 3$ in the limit $m_\pi\rightarrow 0$ .  
Inclusion of the $3\pi$ continuum \eqref{3pi_cont} 
is still likely to underestimate the total spectral function since more 
complicated models of the spectral function involve resonance enhancement of this $3\pi$ continuum \cite{3pi}.
If this limiting form is used up to a cutoff
of $1\,{\rm GeV}$, then the resulting H\"older inequality quark mass bounds are {\em raised} by approximately 10\%, and  
a 14\% effect is observed if the cutoff is moved to infinity.\footnote{The exponential suppression of the large-$t$ region
in the Laplace sum-rule
\protect\eqref{basic_sr} minimizes any errors in this region from this approximation to the  $3\pi$ continuum, and also leads to the observed small difference in extending the cutoff to infinity.}  
Working with the full form \eqref{3pi_cont} complicates the numerical analysis, but the  following simple form (with $t$ in GeV units) is easily verified to be a bound on the $3\pi$ continuum in the region below $1\,{\rm GeV}$.
\begin{equation}
\rho_{3\pi}(t)\ge\frac{4}{3}\left[\left(\sqrt{t}-m_\pi\right)^2-4m_\pi^2 \right]
\label{rho_ap}
\end{equation}
This approximate form of the $3\pi$ continuum again raises the resulting quark mass bounds by approximately 10\%.

\bigskip
This paper is dedicated to the memory of Roger Migneron.
Many thanks to Vic Elias, Gerry McKeon, and Voldoya Miransky for their efforts in organizing  MRST 2001, which resulted in an enjoyable and interesting conference.

\end{document}